\documentclass[sn-basic]{sn-jnl}

\jyear{2023}%

\begin{document}

\title[latent dynamics large networks]{Fast inference of latent space dynamics in huge relational event networks}

\author*[1]{\fnm{Igor} \sur{Artico}}\email{igor.artico@usi.ch}
\author[1]{\fnm{Ernst} \sur{Wit}}\email{ernst.jan.camiel.wit@usi.ch}

\affil*[1]{\orgdiv{Institute of Computing}, \orgname{Universit\`a della Svizzera italiana}, \orgaddress{\street{Via la Santa 1}, \city{Lugano}, \postcode{6900}, \country{Switzerland}}}

\abstract{

Relational events are a type of social interactions, that sometimes are referred to as dynamic networks. Its dynamics typically depends on emerging patterns, so-called endogenous variables, or external forces, referred to as exogenous variables. Comprehensive information on the actors in the network, especially for huge networks, is rare, however.  A latent space approach in network analysis has been a popular way to account for unmeasured covariates that are driving network configurations. Bayesian and EM-type algorithms have been proposed for inferring the latent space, but both the sheer size many social network applications as well as the dynamic nature of the process, and therefore the latent space, make computations prohibitively expensive.  
In this work we propose a likelihood-based algorithm that can deal with huge relational event networks. We propose a hierarchical strategy for inferring network community dynamics embedded into an interpretable latent space. Node dynamics are described by smooth spline processes. To make the framework feasible for large networks we borrow from machine learning optimization methodology. Model-based clustering is carried out via a convex clustering penalization, encouraging shared trajectories for ease of interpretation. We propose a model-based approach for separating macro-microstructures and perform a hierarchical analysis within successive hierarchies. The method can fit millions of nodes on a public Colab GPU in a few minutes. The code and a tutorial are available in a Github repository.
}


\maketitle
\section{Introduction} \label{introduction}

Networks appear in many contexts. Examples include gene regulatory networks \citep{signorelli2016neat}, financial networks \citep{cook2014global}, psychopathological symptom networks \citep{de2017investigation}, political collaboration networks \citep{signorelli2018penalized}, and contagion networks \citep{uvzupyte2020test}. Studying networks is important for  understanding complex relationships and interactions between the components of the system. The analysis can be difficult due to the many endogenous and exogenous factors that may play a role in the constitution of a network. The aim of statistical modelling in this context is to describe the underlying generative process in order to assist in identifying drivers of these complex interactions. These models can assist in learning certain features of the process, filtering noise from the data, thereby making interpretation possible.

In this manuscript we are considering temporal random networks, whereby  nodes make instantaneous time-stamped directed or undirected connections. Examples are email exchanges, bank loans, phone calls, article citations. A common approach to these networks has been flattening the time variable and studying the resulting static network. Although this method simplifies the complexity of the calculations, clearly there is a loss of information about the temporal structure of the process. Most networks are  inherently dynamic. Subjects repeatedly create ties through time. Since the adjustment of ties is influenced by the existence and non-existence of other ties, the network is both the dependent and the explanatory variable in this process \citep{brandes2009networks}. 
Thus rather than viewing this as a static network, we consider the generative process as a  network structure in which the actors interact with each other through the time. Edges are defined as  instantaneous events. This quantitative framework is known as \emph{relational event modelling}.

The basic form of a relational event model as an event history model can be found in  \cite{butts20084} with an application to the communications during the World Trade Center disaster. The model has been extended by \cite{brandes2009networks} to weighted networks: nodes involved in these events are actors, such as countries, international organizations or ethnic groups. An event is assigned a positive or negative weight depending on a cooperative or hostile type of interaction, respectively. 
Other examples of relational event modelling include the work by \cite{vu2017relational} on interhospital patient transfers within a regional community of health care organizations or the analysis of  social interaction between animals \citep{tranmer2015using}. 

In a relational event model the connectivity may depend on the past evolution of the network. Keeping track of the past is challenging for dynamic networks because of the high number of possible configurations (k-stars, k-triangles, etc.) that could be taken into account, as well as their closure time and the time they keep affecting future configurations. 
We thus propose to take some kind of summary of the past configurations. A solution that can both summarize the process and approximate effectively the past information is the idea of a dynamic latent space.  
To describe the latent structure of a network one can think of placing the vertices in a space where the distance between two points describes the tendency or lack of tendency to connect. Among social scientists this is typically called a \emph{social space} where actors with more interactions are close together and vice versa \citep{bourdieu1989social}.
The locations are allowed to change in time. At each time point new connections are formed and  the subjects develop attraction/repulsion that force them to change their social space configuration.
The new configuration is the one that best reflect the new connectivity behavior.
As a result one location at a certain time reflects past information, within the limits of the latent space formulation.
This evolution describes the social history of the subjects, their preferences, and the groups they might join or leave.

The problem of tracking latent locations has been studied by many authors, specifically for the static case, i.e., tracking locations under the assumption that they are fixed over time. For static binary random graphs \cite{hoff2002latent} provide a framework for inference. Some extensions of that model have been developed to overcome the limitations of the latent space formulation  \citep{hoff2005bilinear, hoff2008modeling, hoff2009multiplicative}. The well-known stochastic block model  describes the similarity between the actors by grouping them together, which is similar to latent space formulation. An extension  of stochastic block modelling to relational event data is provided by \cite{dubois2013stochastic}.

An approach for modelling latent space dynamic binary networks was proposed by \cite{sarkar2005dynamic}. The method is based on an initial preprocessing phase where rough location guesses are found through generalized multidimensional scaling, followed by an estimation phase in which the dynamic locations are treated as fixed parameters and optimized via a conjugate gradient method. The distances between nodes are approximated by thresholding larger ones and including an additional penalty for forcing distant nodes to be closer. 

\cite{sewell2015latent} developed a dynamic latent space with node specific parameters that regulate the incoming and outgoing links. Inference is performed via the Metropolis Hastings algorithm and case-control sampling \citep{raftery2012fast} for making it scalable on large  data.  \cite{durante2016locally} developed a Bayesian approach using Polya-Gamma data augmentation for binary connections and Kalman-within-Gibbs sequential learning of Gaussian processes for node dynamics. \cite{artico2022dynamic} tackle the problem from a frequentist perspective where Gaussian processes are estimated via a Kalman-within-EM Relational Event Model that does not require data augmentation and provides a more robust convergence.

\subsection*{The methodology presented}
The aim of this manuscript is to develop an efficient inference scheme  for latent dynamic processes underlying an extremely high dimensional relational event process. The framework is very general and can be extended to networks with weighted edges of any exponential family distribution. There are two dual representations of the process, either as a continuous time exponential or as discrete Poisson counts. Depending on the sparsity of the observed process, one or the other can be selected in the inference procedure. Interpretation of the huge dynamic latent space is made possible thanks to a clustering component that groups nodes with shared trajectories. The inference is performed under the stochastic variational inference framework, where the marginal lowerbound is directly maximized via parallel computing. 

In Section 2 we propose the structure of the latent space and the relational event modeling background with the dual representation of the process. In Section 3 we present the penalized likelihood approach and stress the convex clustering penalization. Section 4 is dedicated to the optimization methodology. We consider a mini-batch stochastic gradient descent, a popular neural network optimization framework, and adapt it for graph data. The algorithm works on sub-sampling the data, hence particular care is given to sparse information handling. In Section 5 we leverage a variational approach to fit jointly both the model parameters and hyperparameters, such as smoothness and clustering. In Section 6 we show that the model can be run repeatedly within the detected clusters to fit a nested latent space. In Section 7 we present a simulation study. Section 8 is an application of our model to the complete Wikipedia history of edited pages. A tutorial is available in our GitHub repository with the code and step-by-step guidance.

\section{Latent space relational event models}
\label{sec:models}

In this section we introduce a general version of a latent space relational event model (REM). We consider a set of actors, defined as a finite vertex set $V=\{1,\ldots,p\}$, that can exchange links or edges in time. In principle, we will consider the exchange of relational events, such as discrete interaction, e.g., sending an email or citing a patent, but one can also consider extensions to the quantitative exchanges, such as import and export. As drivers of the exchange process we consider both endogenous, such as reciprocity, and exogenous variables, such as vertex characteristics. One particular exogenous variable is the relative location of the vertices in some similarity latent space, which itself is defined as a dynamic process. 

We consider a non-homogeneous multivariate Poisson counting process $N= \{N_{ij}(t) \mid i, j \in V, t\in [0,T]\}$ and a smooth process $Z=\{Z_i(t)\in\mathbb{R}^d \mid t \in [0,T], i=1,\ldots, p\}$ relative to some standard filtration ${\mathcal F}$. In particular, we consider $\mathcal{F}$-measurable rate functions $\lambda_{ij}(t)$ that drive the components of the counting process. In particular, we assume that the rates $\lambda_{ij}(t)$ are functions of the underlying positions $Z_i(t)$ and $Z_j(t)$, besides possible other features. 
The features can be of various types: \emph{exogenous} $x_{ij}(t)$,
such as global covariates, node covariates, edge covariates, as well as  \emph{endogenous} ${\mathcal F}_t$-measurable $s_{ij}(t)$, 
where network statistics capture endogenous quantities such as popularity, reciprocity, and triadic closure. The parameter vector $\beta(t)=(\beta_0(t), \beta_1(t))$ determines the relative importance of the various effects. The rate function between nodes $i$ and $j$ at time $t$ is assumed to be
\begin{equation}
	\label{eqn:lambdaC}
	\log\lambda_{ij}(t) = m(z_i(t), z_j(t)) + \beta_0(t)^t x_{ij}(t) + \beta_1(t)^t s_{ij}(t) 
\end{equation}
where $m(z_i(t), z_j(t))$ is a similarity measure between node specific  latent variables. The dynamics are assumed to follow a spline process 
\begin{eqnarray}
\label{eq:splinewalk} 
z_i(t) &=& b(t)^t \alpha^z_i \qquad i=1, \ldots, p\\
\beta(t) &=& b(t)^t \alpha^{\beta},
\end{eqnarray}
for some $m$ dimensional vector of basis functions $b(t)$. $\alpha^z_i$ is the $m\times d$ parameter matrix for a $d$-dimensional spline. The basis type taken to be P-splines as a cheap representation of a Gaussian process. Node specific splines correspond to $z_i(t)$ while $\beta(t)$ are splines shared by all nodes. The similarity measure $m(z_i(t), z_j(t))$ can be $z_i(t)^t\Lambda z_j(t)$ or $-\| z_i(t) - z_j(t) \|^2$. The measure $z_i(t)^t\Lambda z_j(t)$ comes from Hoff's eigen model \citep{hoff2008modeling}. This measure can model multiple similarity forms:  
\begin{itemize}
    \item $\Lambda$ is a $k\times k$ matrix and $\|z_i(t)\|_2=1$: hyper-cube latent space ,i.e., a stochastic block model.
    \item $\Lambda$ scalar and $\|z_i(t)\|^2_2=1$: hyper-sphere latent space where the distance measure is the angle between two nodes. This measure can be approximated locally by the Euclidian distance.
    \item $\Lambda$ scalar: latent space where the inner product defines the degree of similarity between two nodes. This model also express blockmodeling effects embedded into a similarity space. This measure finds interpretation in the angle between two points as a distance, whereas the norm of the single node describes the subjective tendency to make connection.
\end{itemize}

The first two measures, as well as the Euclidian distance, identify a non convex optimization problem while the last one is convex. Although using a convex measure is appealing for the theoretical convergence guaranteed, it suffers from high dimensional saddle points which turn, from a practical perspective, to be similar to a non-convex optimization problem.

We assume a nested latent space, i.e., nodes form communities with common trajectories. These communities can be decomposed into sub-communities that have shared movements within the mother community. This can be repeated for many levels with a progression from the macro scale to the micro scale. We do not make any specific assumption on the shape of these clusters. For most of this manuscript we focus on detecting only the macro cluster level, while in Section 6 we describe the extension the nested levels.

Given the joint formulation  $(Z,N)$ of the state-space and interaction process, we will assume that only the interaction process $N$ is observed and the main aim of this paper is to infer the structure of the smooth process $Z$ and the rate functions $\lambda$, or more specifically, the parameters $\alpha$ associated with their functional form.  
We will consider two cases of the interacting point process defined above. First we consider the general case, in which the relational events are observed in continuous time. This is the traditional setting for relational events. We will also define a relational event model where the interactions can only happen at specific times. For example, bibliometric citations or patent citations only happen at prespecified publication dates. Furthermore, this model allows a generalization to non-binary relational events, such as export between countries, that can be dealt with in the same inferential framework.  
 
\subsection{Continuous time relational event process $N$}
\label{sec:continuousN}

We consider a sequence of $n$ relational events, $E_{\text{cont}}=\{(i_k,j_k,t_k) \mid t_k\in [0,T], ~i_k,j_k \in V, k=1, \dots n\}$ observed according to the above defined relational counting process $N$. 
Conditional on the smooth process $Z$, the distribution of the 
interarrival time for interaction $i\rightarrow j$ is a generalized exponential, with instantaneous rates as described in (\ref{eqn:lambdaC}). The conditional log-likelihood of the process $Z \mid N$
\begin{equation}
	\label{eqn:lik-contN}
	\begin{split}
		\ell(\alpha) =& \sum_{i,j} \left[ \sum_{t \in E_{\text{cont}}(i,j)}\log \lambda_{i,j}(t) \right] - \int_0^T \lambda_{i,j}(t) dt 
	\end{split}
\end{equation}
where the generalized exponential formulation is the one adopted by \cite{rastelli2021continuous}. This likelihood is commonly simplified in the REM literature with the partial likelihood \citep{ perry2013point} relative to the equivalent Cox process \citep{cox1972regression}.

\subsection{Discrete time relational event process $Y$}
\label{sec:discreteN}

Often relational events are ``published'' only on prespecified discrete event times $\mathcal{T}= \{t_1, \ldots, t_{n}\}$. We consider a sequence of $n$ relational events, $E_{\text{disc}}=\{y_{k,ij} ~\mid~ t_k\in [0,T], ~i_k,j_k \in V, k=1, \dots n\}$ where the interactions $i\rightarrow j$ are collected at $t_{k+1}$ from the observation intervals $(t_{k}, t_{k+1}]$, with resulting interval counts $$y_{k,ij}=N_{ij}(t_{k+1})-N_{ij}(t_{k}).$$

We assume that the rate $\lambda$ is constant with respect to the endogenous and exogenous variables inside the collection intervals $(t_{k},t_{k+1}]$. In fact, with respect to the endogenous variable $N$ it makes sense that no further information between the publication dates affects the rates. In other words we  assume that the log link at equation (\ref{eqn:lambdaC}) for the hazard is conditioned to the past information up to time $t_{k}$.

The interval counts $y_{k,ij}$ of the number of interactions between $i$ and $j$ are Poisson distributed with interval rate,
\begin{equation}
\label{eqn:mu_discrete}
\int_{t_{k}}^{t_{k+1}} \lambda_{ij}(t) ~dt=  \lambda_{ij}(t_k)\Delta t_k,
\end{equation} 
where $\Delta t_k = t_{k+1}-t_{k}$.
An advantage of using discrete time is the reduction of the model complexity. In certain real-world processes it is not uncommon to observe thousands, even million of links. A discrete time representation reduces the computational complexity from the number of links to the number of collection intervals.

Given the complete observations $(Z,Y)$, the complete log-likelihood for the discrete time latent space model is
\begin{equation}
 	\label{eqn:lik-discN}
 	\begin{split}
 		\ell(\alpha) =& \sum_{k,i \neq j} -\lambda_{i,j}(t_k) \Delta t_k + y_{k, ij} \log \lambda_{i,j}(t_k)\Delta t_k
 	\end{split}
\end{equation}
Similar to \cite{perry2013point}, who focus on non-homogeneous exponential waiting times, this  approach focuses on non-homogeneous Poisson counts.

This approach can be further generalized to any exponential family \citep{artico2022dynamic} or the zero inflated exponential family \citep{ sewell2016latent}.

\section{A penalized likelihood approach}
\label{sec:penLik}
For inferring the above model we aim to maximize the following penalized log likelihood
\begin{equation}
\label{eqn:loss}
 \ell^P(\alpha) = \ell(\alpha) + P_{\text{smooth}}(\alpha) +  P_{\text{clust}}(\alpha)
\end{equation}
where $\ell(\lambda)$ is either (\ref{eqn:lik-contN}) or (\ref{eqn:lik-discN}) depending on the case. $P_{\text{smooth}}$ is a smoothness penalty on the spline process, $P_{\text{clust}}$ a convex clustering penalty for forcing nodes to be closer. Although in the classic formulation of generalized additive models \citep{wood2006generalized} the the process smoothness is regulated by penalizing the second derivative $\int_0^T  \alpha^2 b(t)'' dt$, for dynamic systems it is more important to consider the first derivative as it regulates the difference  between a static or a dynamic model. Moreover the latent space is not identifiable due to rotations: the resulting dynamics are hence the original nodes trajectories plus infinite infra-time rotations. A first derivative penalty reduces rotations that are misinterpreted as node dynamics. For P-splines the penalty has a convenient form $$ P_{\text{smooth}}(\alpha) =  - \gamma_{\text{smooth}}\sum_{i=1}^p\sum_{k=2}^m \left\|\alpha_{i,k} -\alpha_{i,k-1}\right\|^2$$ with the first order differences on the basis heights. P-splines are a low rank, smooth representation of a Gaussian process. The basis captures the local temporal structure of the process and a finer granularity can be achieved by increasing the number of basis $m$. For $m=n$ we obtain a Gaussian process. Taking $m<n$ has both computational benefit and a potential overfitting reduction.


\subsection{Convex clustering penalty for community detection}
A common problem that arises when using large dimensional models  is that results are dense. It is hard to interpret large amount of parameters. Therefore we simplify our model fit by grouping together nodes into communities that share common movements. It is often more sensible to spot common movements across different nodes in order to separate them from nodes with independent trajectories. We cluster node trajectories with the popular convex clustering penalty \citep{pelckmans2005convex, hocking2011clusterpath, chen2015convex, weylandt2020dynamic}
$$P_{\text{clust}}(\alpha) = -\gamma_{\text{clust}} \sum_i \int_0^T ( z_i(t) - c_i(t) )^2 dt - \gamma_{\text{dist}} \sum_{i<j} w_{ij} \int_0^T (c_i(t) - c_j(t) )^2 dt.$$ Similarly to the smoothness penalty, this penalty finds a discrete simplification 
\begin{align}
\label{eqn:convexclust}
P_{\text{clust}}(\alpha) &= -\gamma_{\text{clust}}\sum_i \| \alpha_i^z - c_i \|^2 - \gamma_{\text{dist}} \sum_{i<j} w_{ij} \| c_i - c_j \|^2 \\*
 &\text{where} \qquad  w_{ij} =  \mathbb{I}_{[0, \gamma_{\text{radius}}]}(\| \alpha_i^+ - \alpha_j^+ \|)
\end{align}
thanks to the P-spline low-rank process representation. This penalty yields a unique solution to a combinatorial problem, which is typically non-convex. This formulation \citep{hocking2011clusterpath, sun2021convex} shrinks the closest nodes in a hierarchical sequence. It consists of a  vector of features $\alpha_i^z$ and a vector of auxiliary variables $c_i$ that correspond to node $i$ centroid. $\alpha^+$ are considered a reliable estimate of the true parameters. The first component $\sum_i \| \alpha^z_i - c_i \|^2$ ensures that the centroids are sufficiently close to the respective nodes while the second component $\sum_{i<j} w_{ij} \| c_i - c_j \|^2$ enforces closer centroids to shorten their distance. The parameters $\gamma_{\text{dist}}$ and $\gamma_{\text{clust}}$ regulate the amount of shrinkage for the centroid-centroid and the centroid-$\alpha$ distance respectively. We can group together centroids that are closer than a certain threshold $\epsilon$. Faster convergence and different cluster shape can be achieved by altering the kernel $w_{ij}$. The kernel aims to increment the penalty locally, its radius is regulated by $\gamma_{\text{radius}}$. Common choices for the kernel are gaussian or discrete, as in (\ref{eqn:convexclust}), whose performances are approximately equivalent.

In the original convex clustering formulation $\alpha$ corresponds to observed features and the kernel is calculated using them as input. A popular attempt of clustering unobserved features comes from \cite{lindsten2011clustering} who clustered the latent states of a Kalman Filter model. Similarly, we estimate these $\alpha^+$ by a pilot optimization phase where we fit the \textit{vanilla} model including the smoothness penalty only. These estimates $\alpha_i^+$ will be  considered as fixed in the further inference. In case we have convexity in both the likelihood similarity measure and in the penalty, we obtain a double-convex optimization problem. The clustering path can be computed by increasing the kernel radius  $\gamma_{\text{radius}}$ or by the shrinkage $\gamma_{\text{dist}}$ in different strategies. As the radius increases, more nodes are included in the kernel and are shrunken, leading to a hierarchical procedure that ends into a single cluster. 

\subsection{A fast convex clustering penalty}
The inclusion of a clustering and distance penalty in the original convex clustering formulation produces, however, a near unidentifiability between $\gamma_{\text{clust}}$ and $\gamma_{\text{dist}}$. Given a fixed radius, multiple combinations of $\gamma_{\text{clust}}$ and $\gamma_{\text{dist}}$ have nearly identical predictive performance without any preference on whether aggregating nodes or not. From a geometrical perspective the amount of shrinkage on $\alpha$ can be held constant for any value of $c$ that follows the path from $c = \alpha$, hence $\gamma_{\text{dist}} = 0$, to the point of centroid  aggregation at $\gamma_{\text{dist}} \rightarrow + \infty$.  
We can bypass the problem by ``dropping" entirely the distance component. The aim is to cluster all the nodes that enter into the kernel. For $\gamma_{\text{dist}} \rightarrow + \infty$ groups of centroids have perfect matching and the minimization of the convex clustering penalty (\ref{eqn:convexclust}) finds analytic solution as
\begin{align}
\label{eqn:conv_clustB}
P_{\text{clust}}(\alpha) &= -\gamma_{\text{clust}}\sum_{i=1}^p   \| \alpha_i - c_i \|^2 \\*
\text{where}\qquad c_i &=  \sum_{j=1}^p \alpha_j \mathbb{I}\{i-j\} /  \sum_{j=1}^p \mathbb{I}\{i-j\}
\end{align}
which has computational complexity linear in $p$ rather than quadratic as before. The value $c_i$ is the average coordinate among all nodes belonging to the same cluster as $i$, which needs to be calculated once for each cluster. $\mathbb{I}\{i-j\}$ simply indicates the cluster assignment or, more precisely, if there exist a path of kernels that connects $i$ to $j$. Thus $\mathbb{I}\{i-j\}$ indicates that $i$ and $j$ belong to the same connected component in the graph constructed by kernel $w_{ij}$. This can be done by updating the kernel adjacency list as the sequence of samples $B$ is filtered by the kernel $w_{ij}$. This implies that not all the pairwise relationships $w_{ij}$ need to be observed, just the ones that relate a node to at least one other node of the same cluster. 

Convex clustering can be considered as a hard clustering method where nodes with unique dynamics are modeled independently, instead of being considered as outliers or abusively allocated to the closest cluster. An alternative approach is proposed by \cite{handcock2007model} with a finite Gaussian mixture model, which may suffer from local minima or high dimensionality. Furthermore, the latter can only detect circular clusters, while in our method we do not specify the cluster distribution.

Alternatively to the kernel aggregation a useful heuristic exists. The fast convex clustering penalty (\ref{eqn:conv_clustB}) can be seen as the analytic equivalent to the \textit{hdbscan} heuristic \citep{schubert2017dbscan} where nodes belonging to the same discrete kernel are sequentially aggregated as the kernel enlarges. This heuristic can suggest good candidate radii to test and offer a more robust allocation. Moreover the $\gamma_{\text{dist}}\rightarrow \infty$ convex clustering version can be interpreted in a more general perspective where any clustering or aggregation algorithm can be used and the resulting cluster allocation can be plugged in the model. Thus our approach opens the door to a supervised clustering selection method for a wide range of existing algorithms.

\section{Optimization}
\label{sec:inference}
The computational complexity for optimizing the model described in section (\ref{sec:penLik}) is prohibitive when the data dimension is very large. In these cases it is necessary to restrict the inference over subsamples of the data. 
A method that we borrow from machine learning is the so called mini-batch gradient descent. It consists into taking random subsample from the data  named mini-batch $B$, where $B \subset E$ and $E = E_{\text{cont}}$ or $E = E_{\text{disc}}$, according to the case. The mini-batch  has typically small size $n_b =  \mid B \mid $. The fast computation, mostly matrix operations, is restricted to the mini-batch. Over this subset the likelihood $\ell(\alpha)_B$ is calculated and a gradient step is taken, such $\alpha \leftarrow \alpha + \psi \nabla \ell(\alpha)_B$. The procedure is repeated, sampling new mini-batches B,  until convergence. As a result of the subsampling the gradient is an unbiased estimator of the full gradient. The mini-batch gradient trades variance for computational and memory cost. For a certain mini-batch size, stochastic gradient descent reaches the minimum faster than a deterministic gradient. The gradient update step is a Newton step where the costly second derivative matrix is substituted by a cheap but unknown $\psi$ parameter. As a result, the missing Hessian leads to the gradient elements having wrong individual scale, hence wrong global direction in the gradient vector.
The past literature, e.g. \citep{ruder2016overview, duchi2011adaptive}, has focused on two main issues: decreasing the gradient variance and rescaling the gradient estimate.
Both problems are solved by the popular \textit{Adam} \citep{kingma2014adam}. In \textit{Adam} the gradient update is formulated as a state-space model, where the gradient moments are thought as latent states. Leveraging a simple, univariate form of the Kalman Filter, known as Exponentially Weighted Moving Average (EWMA), the update has the form
\begin{eqnarray*}
g &\leftarrow& \nabla \ell(\alpha)_B \\
m_{k} &\leftarrow& \xi_1 m_{k-1} + (1-\xi_1) g \\
v_{k} &\leftarrow& \xi_2 v_{k-1} + (1-\xi_2) g^2 \\
\alpha_k &\leftarrow& \alpha_{k-1} +  \psi  \frac{m_k}{v_k}
\end{eqnarray*}
at iteration $k$, $m_{k}$ and $v_{k}$ are the gradient first and second moments, respectively. Hence the moments are a weighted average with the past moments, where the weights decrease exponentially in time. The  $\xi$ parameters regulate how much of the past information is used to update the current moments.
Thus \textit{Adam} provides an estimator for the first two gradient moments. The benefit from the averaging is the variance reduction of these moments, although some bias might be introduced if the process relies too much on the past. Moreover, leveraging the Bartlett identity $E[\frac{\partial^2}{\partial^2 \alpha} \ell(\alpha)] = E[(\frac{\partial}{\partial \alpha} \ell(\alpha))^2]$ we have that $v_{k}$ is an estimator of the diagonal elements of the Hessian matrix.  Imposing locally, i.e. at iteration $k$, the assumption of a spherical covariance matrix between the parameters, the inverse of the diagonal Hessian applies an effective rescaling to the gradient elements. The algorithm can also tackle high parameter correlation or ridge problems by learning the correct direction from the past steps. The lack of the off-diagonal Hessian elements is hence replaced by the gradient averaging over the past noisy directions.
The optimization is performed until the algorithm reaches the maximum or, more precisely, a stationary distribution at the maximum. This stationary distribution has been extensively studied and in some cases it can be considered as a  posterior distribution \citep{mandt2017stochastic}. The optimization is stopped if the algorithm does not find a new maximum after a reasonably high number of iterations. 

Although \textit{Adam} has shown to be effective in many scenarios, it has some side effects. 
The algorithm can suffer from pathological cases of severe parameter scale imbalance or large gradients variance (see Section \ref{sec:spGrad} about sparsity).
The problem of scale is commonly tackled in machine learning via parameter normalization. In our case it can be mitigated by using  basis splines which share similar scale in the weights, such as P-splines.

\subsection{A sparse gradient update problem}
\label{sec:spGrad}
When working with high dimensional problems, the amount of information contained in the mini-batch determines the success of the optimization. In our model the shortage of information corresponds to the problem of sparsity. In this section we tackle two types of sparsity: sparsity in the sampled connectivity and sparsity in the sampled parameters. \textit{Adam}, by increasing the long term memory parameters $\xi$, is designed for solving sparse update problems. However in extremely sparse scenarios the gradient variance can become too high and the EWMA cannot recover a decent signal from the noise.

\subsubsection{Sparsity in the parameters}
The mini-batch size determines how many nodes and time points, hence parameters $\alpha_i$, are included in the current iteration. The gradient over the missing parameters is zero, therefore the EWMA performs a smooth averaging over a sparse vector. A way for reducing the gradient variability is to include as many parameters as possible in the mini-batch. A mini-batch of size $n_b$ on average contains $0.632 \times 2n_b$ nodes, where 0.632 is the resampling bootstrap ratio. Given the local structure of P-spline basis, every time point corresponds to 4 non zero basis. We hence update an average of $0.632 \times 2n_b \times 4 \times d$ parameters over a total of $pmd$ parameters. Fixing $m=10$ allows to fit a 10 degrees of freedom function, a value that is sufficiently high in most applications. The gradient is sufficiently dense as long as the ratio $0.632 \times 8n_b/pm$ is close to 1. The size of the mini-batch should hence grow linearly with the nodes $p$. Possible choices are between $n_b = p$  and $n_b = 2p$ for a ratio of approximately 0.5 and 1 respectively.
These values correspond to a sparsity level that \textit{Adam} can handle easily, see Figure \ref{fig:vary_batch}. 
Moreover, the calculations are made under the worst case scenario where all the degrees of freedom are necessary. In case the effective degrees of freedom is less than $m$ the smoothness penalty defines a dependency chain  over the basis parameters, i.e.,  parameters are more correlated and move together. The level of smoothness regulates how local is this kind of dependency: the higher the smoothness, the lower the effective number of parameters.

A similar reasoning applies to centroids $c_i$. The simplification in  (\ref{eqn:conv_clustB}) solves another important sparsity problem. If we were using the original penalty (\ref{eqn:convexclust}) the quadratic cost of the distance component would require some sort of subsampling, i.e., a mini-batch penalty $\sum_{i,j \in B} w_{ij} \| c_i - c_j \|^2$. Since the chances of randomly sampling two close nodes are almost zero for large networks, the vast majority of elements would be excluded by the kernel. As a consequence, the level of sparsity of the gradient with respect to $c$ would be even higher than for the splines. This results into an ineffective shrinking of centroids. Instead (\ref{eqn:conv_clustB}) solves the problem by removing this component. The gradient is calculated over all the centroids and they are aggregated by the kernel only. In Appendix \ref{app:mbclust} we propose an alternative mini-batch convex clustering penalty.

\cite{rastelli2021continuous} constructed  the  mini- batch by sampling a set of nodes, rather than edges like our case, including all the dependencies with the remaining nodes. This produces a node-wise update where the information tend to focus too much on the single node and very little on the others. The algorithm needs to cycle over all the nodes before focusing on the same nodes again.
The optimization is carried by a memory-less Stochastic Gradient Descent that cannot compensate for the imbalance. 
This two factors might result into slow or false convergence.

\subsubsection{Sparsity in the links}
Sparsity not only occurs in sampling nodes, but also in the observed data and in the information of the gradient. We refer to this as gradient sparsity in a more general sense.
The problem of independent sampling in a sparse large network is that distant nodes are sampled more often, which do not interact. The large amount of zeros that overcrowd the mini-batch is redundant, hence very little informative.  As a result, the gradient taken over the mini-batch rarely contains  information about the connectivity between two nodes. The redundancy lies in the fact that the macro level structure of a large network can be summarized by few ''compound'' zeros that connect macro components.

Some authors have tried to solve the problem by partitioning the latent space into blocks. Hence the overall number of interactions can grow only linearly with the number of nodes \citep{rastelli2018computationally}.
Case-control sampling overcomes the redundancy in the data by including in the sample as many links as possible (cases), with a minimal inclusion of zeros (controls). The idea consists on dropping the majority of zeros and making few of them representative of the entire non-interacting population. The only consequence of the case-control sampling is the increase of variance in the estimates, but this is commonly compensated by the large amount of data. 
\cite{raftery2012fast} give a detailed procedure on how to perform stratified case-control sampling for static binary networks. Shortest path distances are  used as a proxy of the latent distance, allowing for  stratification of controls  at different lengths. Controls are sampled in each stratum for each node. Particular care must be paid to sampling the same control for the two nodes in order to avoid unnecessary biases in the case-control weights, as the two pushing forces might differ substantially if the two nodes have substantially different centralities. The procedure approximate the likelihood and successfully capture both macro and micro structure in the latent space. However, the preprocessing phase where controls are sampled is both computationally and memory expensive. 

A cheaper solution is proposed in the Supplementary Material of \cite{sewell2016latent}, applied to temporal networks. The stratification is dropped and the controls are sampled at random, capturing mainly the macro structure. An additional control set contains all the non-interactions of nodes with at least one interaction during the time span. Although this set accounts for a minimal micro-structure, its memory requirements can explode easily. The set size indeed increases as time goes to infinity since it is more likely to observe at least one interaction between two nodes. 

In our approach we drop the micro community structure since we have a clustering formulation. We therefore can make a further simplification in the case-control sampling. Sampling controls at random capture mainly the macro structure as you sample more frequently distant nodes. We propose two different model formulations. Depending on the level of sparsity of the process,  a continuous time or a discrete time formulation.

\subsection*{A discrete time model for dense data} The model in  (\ref{eqn:lik-discN}) can be used when the network present many interaction. Clearly storing the adjacency matrix elements (the square of the nodes $\times$ the number of time intervals) is unfeasible for large networks, hence we restrict this usage only for cases where the interactions can be calculated on line. For such cases there is no need of storing all the pairwise interactions as they can be calculated during the sampling phase.

\subsection*{A continuous time model for sparse data}
We propose the case-control version for the inference of a continuous time relational event model (\ref{eqn:lik-contN}). A popular approach in the REM literature \citep{butts20084, brandes2009networks,vu2017relational} is to maximize the so-called partial likelihood  $$ PL(\alpha) = \prod_{tij \in E} \frac{\lambda_{ij}(t)}{\sum_{kl}\lambda_{kl}(t)}$$ of the Cox process $N$ at (\ref{eqn:lik-contN}). As the risk set in the denominator is computationally challenging, \cite{vu2015relational} following \cite{borgan1995methods}  show that a random subset of the risk set yields a consistent estimator for the model parameters. \cite{lerner2020reliability} pushed this concept to the limit by showing that sampling one single control is a sufficient statistic for the risk set, fitting successfully a REM over millions of nodes.
The partial likelihood in that case is
\begin{equation}
\label{eqn:caseControlPartial}
\ell(\alpha) =  \sum_{tij\in E} \log \frac{\lambda_{ij}(t)}{\lambda_{ij}(t) + \lambda_{i^*j^*}(t)}
\end{equation}
where $i^*,j^*$ is a sampled control at time $t$. This case-control sampling hence allows to store in memory only the history of links. The mini-batch $B$ is composed by sampling half links and half controls, where new controls are sampled at each likelihood evaluation. The only drawback of subsampling one single element is the increase of variance in the estimates, as it is inversely proportional to the number of controls subsampled. This is compensated by the vast amount of data that comes from a large network. Similarly to \textit{Adam}, the case-control likelihood trades variance for computational efficiency.

In case the links are dense within communities a case-control discrete time  model is considered in Appendix \ref{app:mixedsparsity}.

\subsection{Mini-batch model}
The calculation of the mini-batch loss should be computed efficiently.
We require that the matrix operations grow linearly with the number of nodes $p$.
At each iteration we sample a mini-batch $B \subset E$, where $E$ is either $E = E_{\text{cont}}$ or $E = E_{\text{disc}}$,  consisting of randomly sampled pairs $i,j$ and time $t$ from the data set. We set the mini-batch size $ \mid B \mid =2p$ to ensure that the gradient is calculated over the majority of parameters. Lower sizes might update only a little portion of nodes, destabilizing the optimization algorithm as discussed in Section \ref{sec:spGrad}. All the matrix operations and gradients are computed over the mini-batch penalized likelihood
\begin{equation}
\label{eqn:mini_loss}
\ell(\alpha)_B^p = \frac{ \mid E \mid }{ \mid B \mid }\ell(\alpha)_B + P_{\text{smooth}}(\alpha) + P_{\text{clust}}(\alpha), 
\end{equation}
where $\ell(y, \lambda)_B$ is the likelihood evaluated over B, given in  (\ref{eqn:caseControlPartial}) or (\ref{eqn:lik-discN}) for sparse or dense network scenarios. Similarly to case-control weights, $\frac{ \mid E \mid }{ \mid B \mid }$ rescales the likelihood component accounting for the downsampling. $P_{\text{smooth}}$ and $P_{\text{clust}}$ do not require any subsampling since they have a computational complexity that is linear in $p$. Additionally they yield a faster optimization as the full parameters dependencies are included.

\section{Stochastic Variational Inference}
In this section we discuss how to estimate both the model parameters $\alpha$ and the hyper parameters $\gamma = (\gamma_{\text{smooth}}, \gamma_{\text{clust}}, \gamma_{\text{radius}})$. Given the full parameter vector $\theta = (\alpha, \gamma)$ a naive choice for maximizing the marginal likelihood $p(y) = \int p(y  \mid  \theta) p(\theta) d\theta$  can be k-fold cross-validation. Validation sets are iteratively removed from the model inference and hyper-parameters are selected as the best performing in these sets. Although cross-validation is a good way for assessing hyper-parameter tuning in dense networks, it can be unreliable for sparse scenarios. 
In order to avoid removing relevant information about single node dynamics, the validation set should be as small as possible. This leads to a high  number of validation sets, hence high computational burden. Moreover, the number of hyper-parameters is recommended to be either low or weakly dependent, which is not our case.

Our proposed approach for maximizing $p(y)$ is via stochastic variational inference  \citep{kingma2013auto, hoffman2013stochastic, blei2017variational, kucukelbir2017automatic}. Variational inference aims to maximize the following lowerbound of the marginal likelihood
\begin{eqnarray}
\label{eq:ELBO}
\notag
\log p(y) &=& \log \int p(y \mid  \theta) p(\theta) d\theta = \log \int p(y \mid  \theta) q_{\mu, \sigma}(\theta) \frac{p(\theta)}{q_{\mu, \sigma}(\theta)} d\theta \\
\notag
&=& \log \mathbb{E}_{q_{\mu, \sigma}}[ p(y \mid  \theta) \frac{p(\theta)}{q_{\mu, \sigma}(\theta)}] \geq \mathbb{E}_{q_{\mu, \sigma}}[ \log p(y \mid  \theta)] + \mathbb{E}_{q_{\mu, \sigma}}[\log p(\theta) - \log q_{\mu, \sigma}(\theta)] \\
&=& \mathbb{E}_{q_{\mu, \sigma}}[ \log p(y \mid  \theta)] - D_{\text{KL}}[q_{\mu, \sigma}  \| p]  = \mathcal{L}(\mu, \sigma)
\end{eqnarray}
where the unknown true density $p(\theta)$ is in practice replaced by an arbitrary prior distribution and the posterior distribution is approximated by the variational density $q_{\mu, \sigma}(\theta)$. A common choice is independent Gaussian $q_{\mu, \sigma}(\theta) = \prod_{i=1}^{p+3} q_{\mu_i, \sigma_i}(\theta_i)$ where all posterior dependencies are ignored and inference reduces to the first two posterior moments $\mu, \sigma^2$. Differently from the mean field approach that aims to find recursive closed form of $q_{\mu, \sigma}(\theta)$, stochastic variational inference aims to direclty maximize (\ref{eq:ELBO}) where the untractable components of the lowerbound are approximated via Monte Carlo integration \citep{kingma2013auto, kucukelbir2017automatic}. All parameters can hence be updated simultaneously using \textit{Adam} stochastic gradient optimization.  The only element that requires Monte Carlo evaluation is the mini-batch likelihood. As shown by \cite{kingma2013auto} in the expectation $\mathbb{E}_{q_{\mu, \sigma}}[ \ell_B(\alpha)] =  \frac{1}{H}\sum_{h=1}^H \ell_B(\alpha^h)$ the number  $H$ of Monte Carlo replicates  drawn from $q_{\mu, \sigma}(\alpha)$ can be reduced to 1 when the mini-batch size is sufficiently large and the optimization is performed via moving average gradient scheme.  
At each iteration we draw one Monte Carlo sample $\alpha^*$ from the variational density  $q_{\mu, \sigma}(\alpha)$ obtaining the mini-batch lowerbound
\begin{align*}
\mathcal{L}_B (\mu, \sigma) &= \frac{ \mid E \mid }{ \mid B \mid }\ell_B(\alpha^*) + \mathbb{E}_{q_{\mu, \sigma}} [P_{\text{smooth}}] + \mathbb{E}_{q_{\mu, \sigma}} [P_{\text{clust}}] - D_{\text{KL}}[q_{\mu, \sigma} \| p] \\*
 &\text{where} \qquad \alpha^* = \mu + \sigma \epsilon \qquad \epsilon \sim N(0,1),
\end{align*}
which is the quantity we maximize. The reparametrization $\alpha^* = \mu + \sigma \epsilon$ ensures that the gradient  is not affected by noise in updating the parameters $\mu, \sigma$. Moreover, we recommend to initialize $\sigma$ small, as the single-sample Monte Carlo integration is prone to diverge for large variance.
In a variational context the two penalties naturally translate into Bayesian priors. The three remaining expectations $D_{\text{KL}}[q_{\mu, \sigma}(\theta)  \| p(\theta)]$, $\mathbb{E}_{q_{\mu, \sigma}} [P_{\text{smoooth}}]$, $\mathbb{E}_{q_{\mu, \sigma}} [P_{\text{clust}}]$ have simple close form solutions thanks to the Gaussianity and independence, see Appendix \ref{app:elbo} for details.  

Variational inference works particularly well in settings where $q_{\mu, \sigma}(\theta)$ provides a sufficiently good approximation of the posterior, i.e., the lower bound reaches a sufficiently close value to the marginal. The independence assumption on $q_{\mu, \sigma}(\theta)$ is appropriate for a posterior that is approximately independent or, like in our case, locally dependent. The conditional dependency induced by observing the data, i.e., the posterior covariance, is locally present for  close nodes and adjacent time points. Hence latent network representations combined with a Gaussian process are particularly suited for variational inference, as it ignores a relatively small amount of information when approximating with an independent posterior. Once again we fit the macro scale by sacrificing the micro scale dependencies. 
Finally, our model can be seen as variational autoencoder \citep{kingma2013auto} with the addition of penalties. Despite its most common usage as image generator, a variational autoencoder is a more general framework for representing any Bayesian inference problem as an encoder-decoder. For our model the decoding side is fully structured by the link function while the encoder reduces to a selector operator that associates an edge to the respective posterior node positions in the latent space.

\section{Marginalization: A hierarchical community model}

For static networks, repeated community detection can be used to detect hierarchies of nodal communities. Our methodology can be seen as a dynamic model-based partitioning of the nodes. 
By repeated application of our method we can obtain nested communities in dynamic networks. The concept of nested communities is appealing to practitioners, where  interpretation is simplified via nested structures. 

This divide-and-conquer approach suits well the model purpose. Given the set of clusters the latent space model is estimated recursively inside each cluster. This nested procedure can be iterated multiple times as long as the variance of the locations allows for a meaningful community discovery. This procedure is performed over clusters of reasonable size: unassigned nodes or small communities are left untouched. 
This fitting procedure can be seen as adding a random effect to the model for explaining within cluster variance.

Under the latent space assumption any marginalization or sub-sampling of the original network is a coherent estimator of the locations and therefore the inference in the micro structure can be done regardless to the macro structure. Given that any subset $V'$ of $V$ maintains the same distances among nodes, the distribution of the restricted node set $P_{V'}$ is the same as the marginalized distribution of the full model $P_V \mid _{V'}$. This invariance means that it is unimportant to which node set the observed nodes actually belong. The model is therefore invariant under marginalization.

The micro communities formulation offers various advantages. In case the community is sufficiently small we can account for all the dependencies with a full covariance matrix for the variational parameters as proposed in \cite{blei2017variational} or a low rank approximation of it backed by importance sampling \citep{zhang2021pathfinder}. Moreover time dynamics can  have a finer granularity, thus they can be captured with a higher number of spline basis or a Gaussian process. The Extended Kalman filter model proposed in \cite{artico2022dynamic} performs sequential learning of Gaussian processes embedded in dynamic networks. The model can be thought as a special case of variational Expectation Maximization where the posterior is approximated by a multivariate Gaussian matching the first two moments.

\begin{figure}[t]
    \centering
    \includegraphics[width=0.7\textwidth, keepaspectratio]{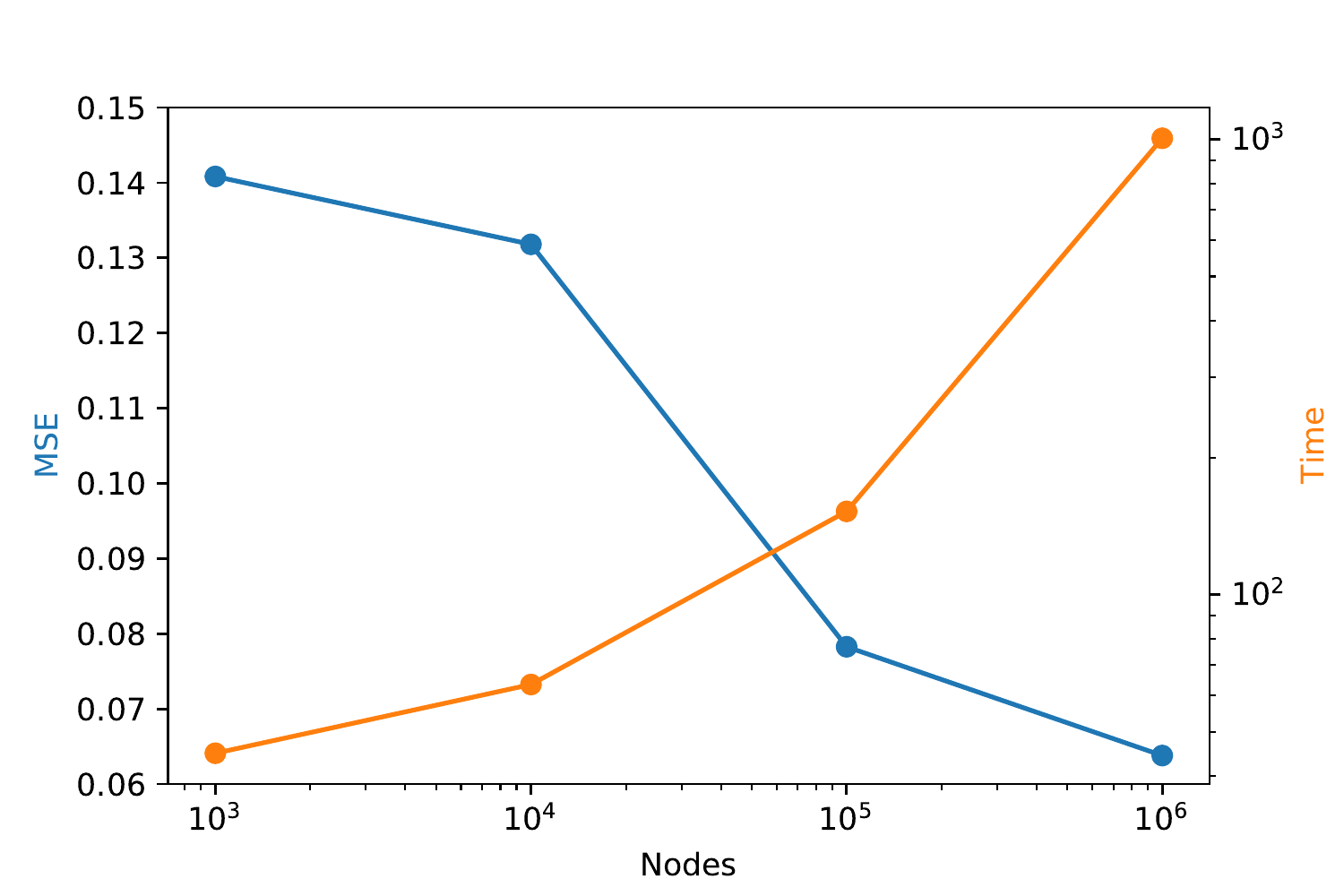}
    \caption{Model average performance and computational time. Goodness of fit (MSE) improves as the nodes increase. Computational time (in seconds) in the log-log plot express a sub-linear increase, showing that the model scales at most linearly with the number of nodes. A network with $10^5$ nodes takes approximately 5 minutes of training while a network with $10^6$ nodes takes approximately 20 minutes.}
    \label{fig:vary_p}
\end{figure}

\section{Simulation study}
We dedicate this section to investigating the features of the estimation procedure.
We are particularly interested into exploring: goodness of fit and computational time as the number of nodes varies, the convergence behavior for different mini-batch sizes, the comparison of different models for different sparsity scenarios, accuracy of classification in different clustering settings. As the locations are not identifiable up to an arbitrary rotation, translation and mirroring, MSE is calculated by pre-processing results via a Procrustes transformation, searching the best rotation and translation that match the truth.
Simulations are repeated 10 times and nodes starting points are set at 0.
\paragraph{Vary number of nodes}
The first scenario is presented in Figure \ref{fig:vary_p} where the average MSE between the fitted and true trajectories is calculated. The goodness of fit improves with the nodes. This support the consistency of the latent location estimator as it converges to the true locations for large number of nodes \citep{shalizi2017consistency}. 

Average computational time, in a log-log plot, follows a sub-linear increase showing that the model scales at most linearly with the number of nodes. The increasing angle indicates the limitations of the GPU used in these analyses. One million nodes indeed requires a significant use of memory, which slow down computations. All our analysis have been conducted with a standard and free Colab GPU, which struggle beyond 4 million nodes.  We suggest switching to more powerful GPUs for larger settings. 

\begin{figure}[t]
    \centering
    \includegraphics[width=0.7\textwidth, keepaspectratio]{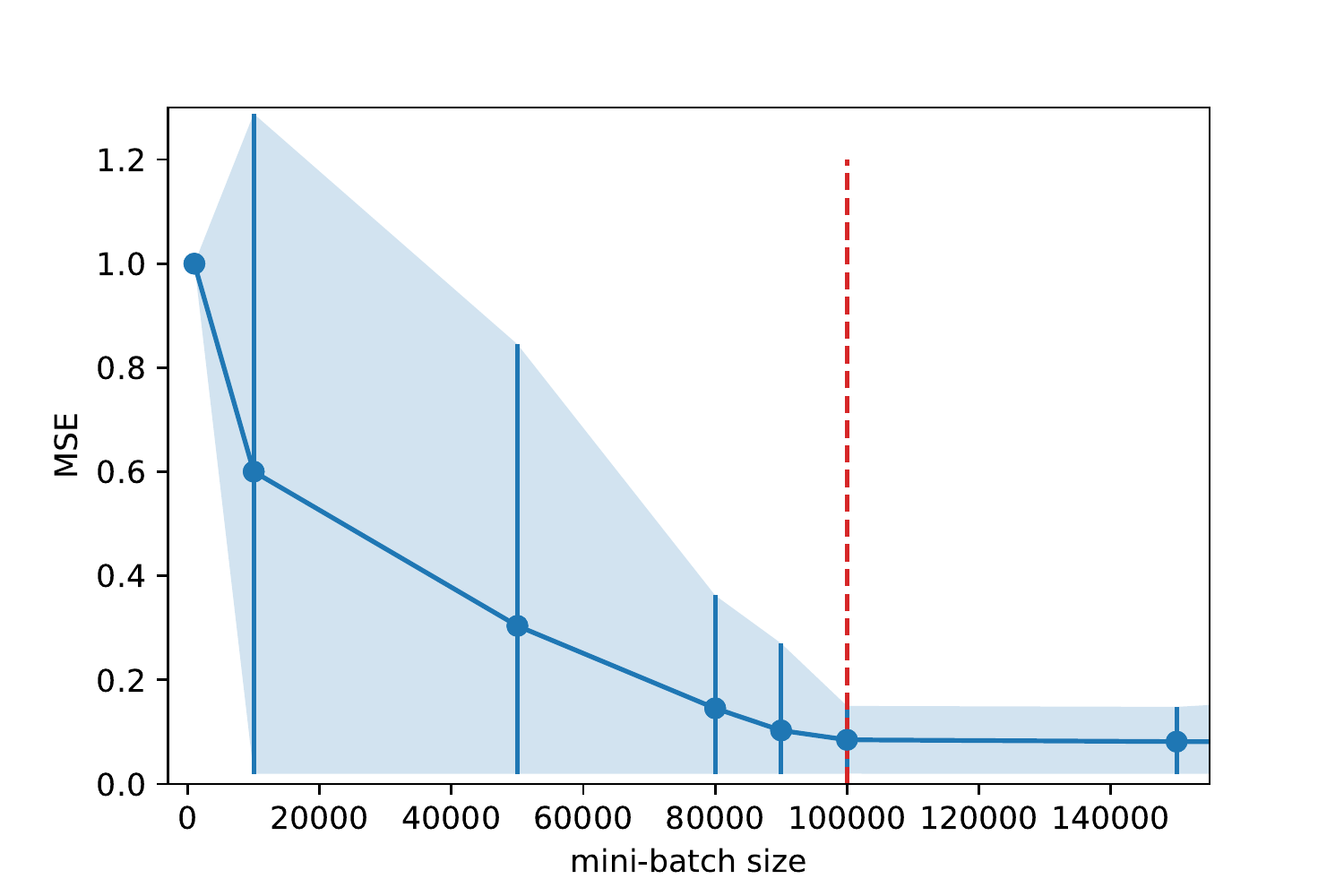}
    \caption{Vary the mini-batch size. MSE (blue line) improves as the mini-batch increases. The high standard deviation (blue bars and shades) highlights false convergence behavior below the safe threshold(vertical red line) where the sparse parameter update is not sufficiently informative. For the lowest mini-batch size the algorithm does not make any meaningful movement from the starting points. }
    \label{fig:vary_batch}
\end{figure}

\paragraph{Vary mini-batch size}
The second set of experiments consists of varying the mini-batch size. We use as standard setting a network with $10^5$ nodes. Figure \ref{fig:vary_batch} shows how a low mini-batch size can cause false convergence as the level of sparsity in the parameter update does not carry enough information for a proper gradient direction recovery, as mentioned in  Section (\ref{sec:spGrad}). For the lowest mini-batch size considered ($10^3$) the fit has both poor MSE and low standard deviation. This means that the algorithm does not move. By increasing the size we have a gradual improvement of the MSE, however the high standard deviation points to a serious instability, it might or not converge to a good value. The behavior stabilizes above a mini-batch size of $10^5$, giving both low MSE and stability. Hence a mini-batch size $n_b = h \times p$, with  $h>1$,  can be considered a safe ratio for ensuring the fitting.

\begin{figure}[t]
    \centering
    \includegraphics[width=0.5\textwidth, keepaspectratio]{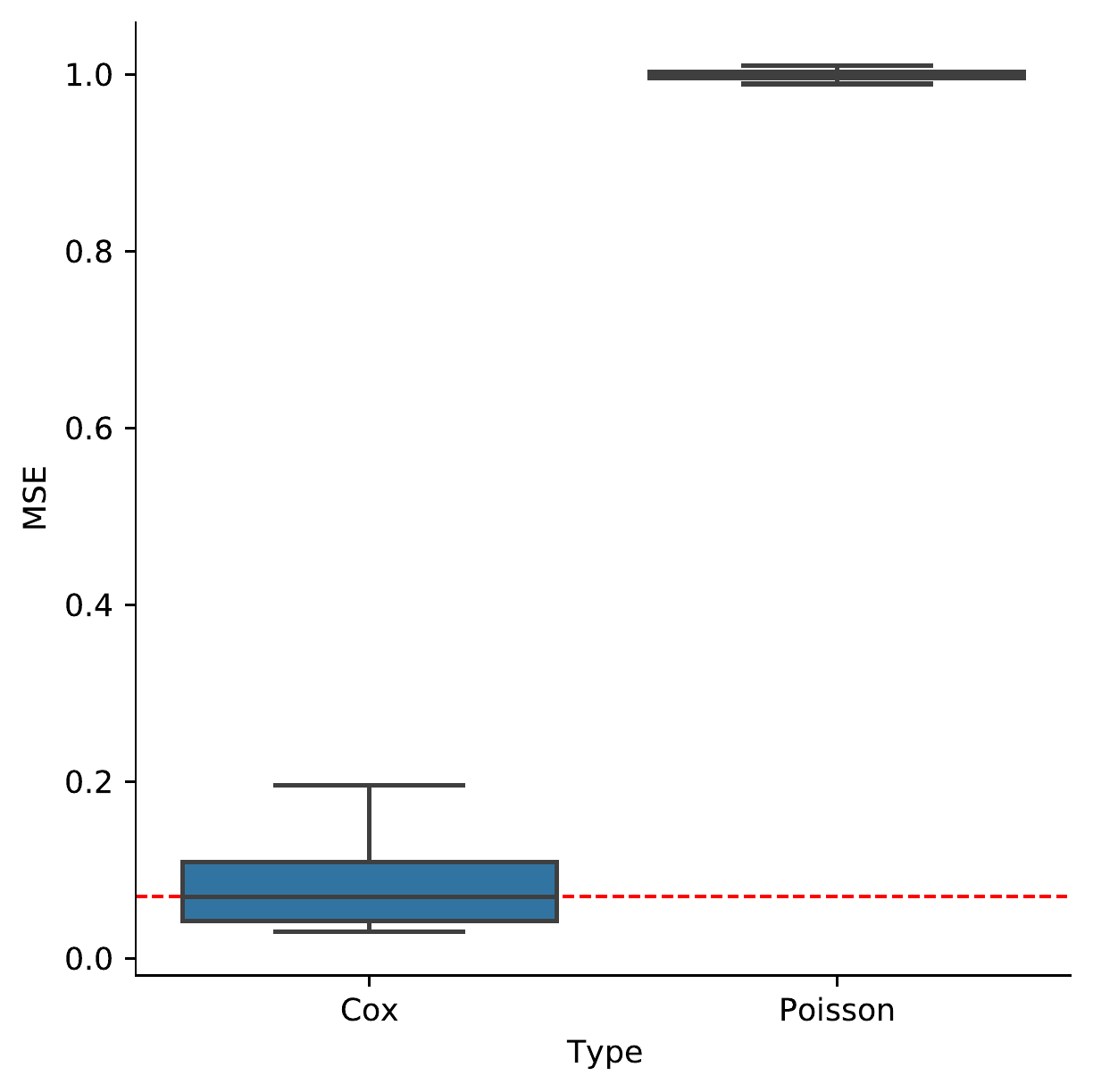}
    \caption{High sparsity scenario. Red line is the Poisson model performance for the dense scenario, we keep it as a benchmark. The Poisson fit deteriorates showing inadequacy for catching sparse behaviors. The Cox model shows to meet the benchmark with comparably fit. }
    \label{fig:vary_sparsity}
\end{figure}

\paragraph{Vary sparsity in the links}
We compare the behavior of the algorithm under different sparsity levels for some models presented in \ref{sec:spGrad}. We compare the Poisson model for dense network activity with the Cox model for the sparse case, showing that they have comparable performance. The Poisson model performs optimally in dense scenarios, however it deteriorates as sparsity increases, in a behavior very similar to Figure \ref{fig:vary_batch}. In Figure \ref{fig:vary_sparsity} we show that the Poisson model in the sparse scenario performs inevitably worse then in the dense scenario. The dense Poisson fit is represented by the dotted red line.
The sparse Cox model presents an MSE very similar to the dense Poisson, showing that the case-control sampling in the risk set does not deteriorate the fit significantly and hence the partial likelihood correctly channel the information necessary for inference.

This case of study underline the relationship between the two types of sparsity mentioned in Section (\ref{sec:spGrad}).
Sparsity, whereas in the parameters or in the data, results into a partial recovery of the true dynamics. By increasing the sparsity we have a worsening as more spline basis parameters never leave the starting point at 0.

\begin{figure}[t]
    \centering
    \includegraphics[width=0.7\textwidth, keepaspectratio]{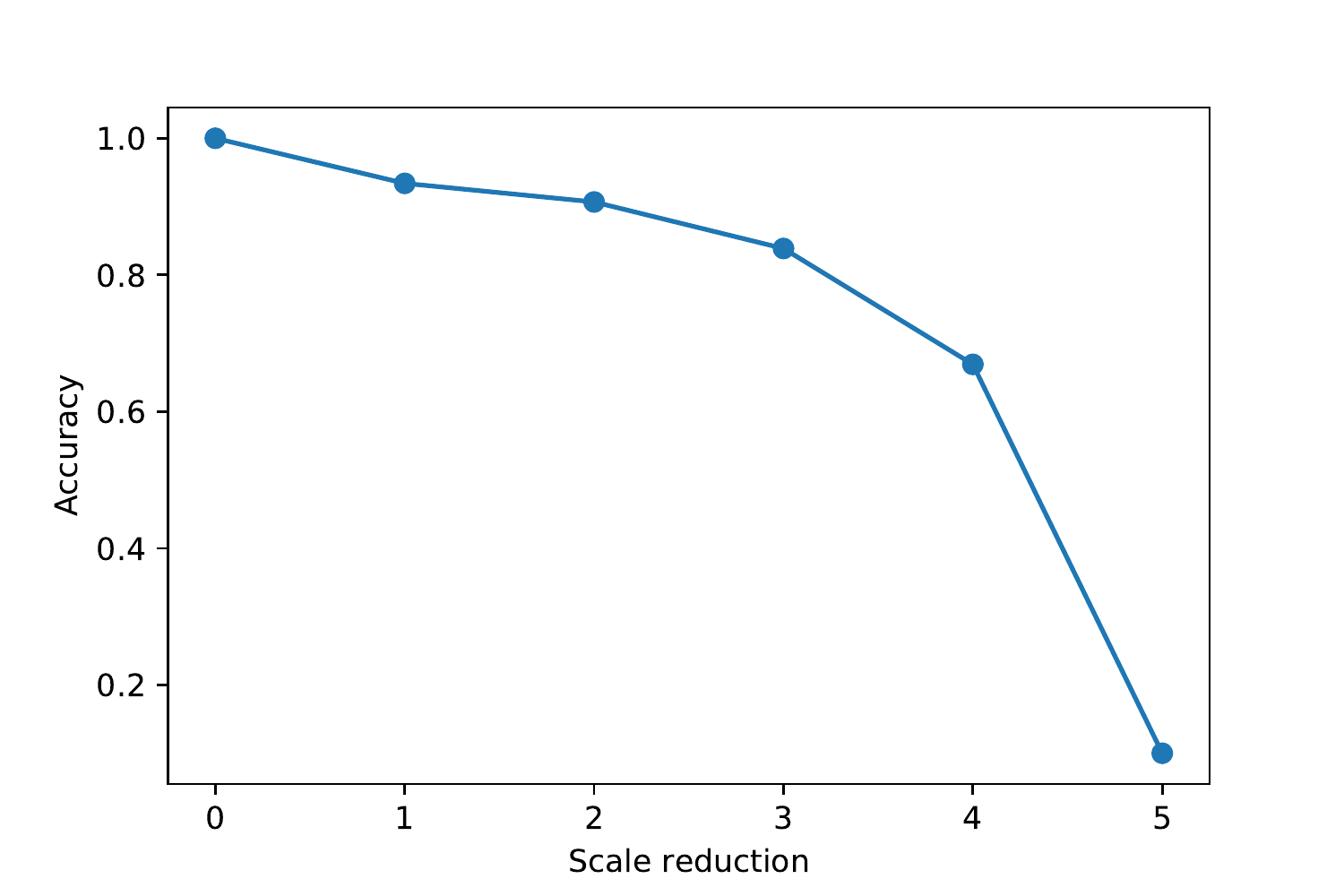}
    \caption{Clustering performance. Perfect allocation for separable clusters. As shrinkage increases, clusters are put closer, letting higher chances for nodes trajectories to overlap. Clustering accuracy deteriorates as a consequence.}
    \label{fig:vary_clustering}
\end{figure}

\paragraph{Vary clusters vicinity}
We conclude  by showing the clustering accuracy as the scale of the synthetic latent space reduces toward 0, letting nodes become closer. In Figure \ref{fig:vary_clustering} we show that the proposed method correctly allocates nodes as long as the space is sufficiently separable. The first point indeed shows perfect classification. The more the nodes are closer, the more the individual node variance becomes influent, the harder the model discriminates between different clusters. 

\section{Data analysis: Wikipedia editing network}
Wikipedia editing history consists of the history of all the editing events of articles by the editors since the foundation of Wikipedia in 2001. This massive bimodal event history data set includes approximately 361 million links, 6.7 million editors, 5.5 million articles and hence $40$ trillion of possible dyadic interactions. The focus is modeling the latent drivers that might explain the user editing behavior. \cite{lerner2020reliability} successfully fitted a Cox proportional hazard model, where endogenous effects such as repetition, two-step reciprocity, individual strength and assortativity are taken under consideration. Their model includes a total of 5 parameters. In this manuscript we propose a more complex form of endogenous effect, the dynamic latent space, where we fit several millions of parameters. The model we propose is the following

\begin{eqnarray}
    \log \lambda_{ij}(t) = -\| z_{i}(t) - z_{j} \|_2^2 + \text{propensity}_i + \text{propensity}_j
\end{eqnarray}

that describes an Euclidean latent space where user $i$ and article $j$ have subjective propensity of editing and being edited respectively.  We chose to keep articles static in time. This improve interpretation as the space becomes a \textit{latent topic space} where editors move when they approach new articles. The interpretation of such space is powerful as certain regions correspond to topics that have a certain degree of relationship, i.e.,  the similarity induced by the heterogeneity of editors background.

The use of the propensity random effect, the Euclidian distance and the static article is justified by a substantial improvement in the model fit. Without these assumptions the model places the editors and articles into two separate clouds, with minimal dynamics. We filtered the data as most of the editors modify few articles only at the beginning of their subscription. We retain editors that have at least 15 interactions. For characterizing the topic space is sufficient to keep the most popular articles, edited at least 100 times. The overall network contains 209.737.058 links, 706.820 articles, 572.586 editors for a total of 1.279.406 nodes.

\begin{figure}[t]
    \centering
    \includegraphics[width=\textwidth,height=\textheight,keepaspectratio]{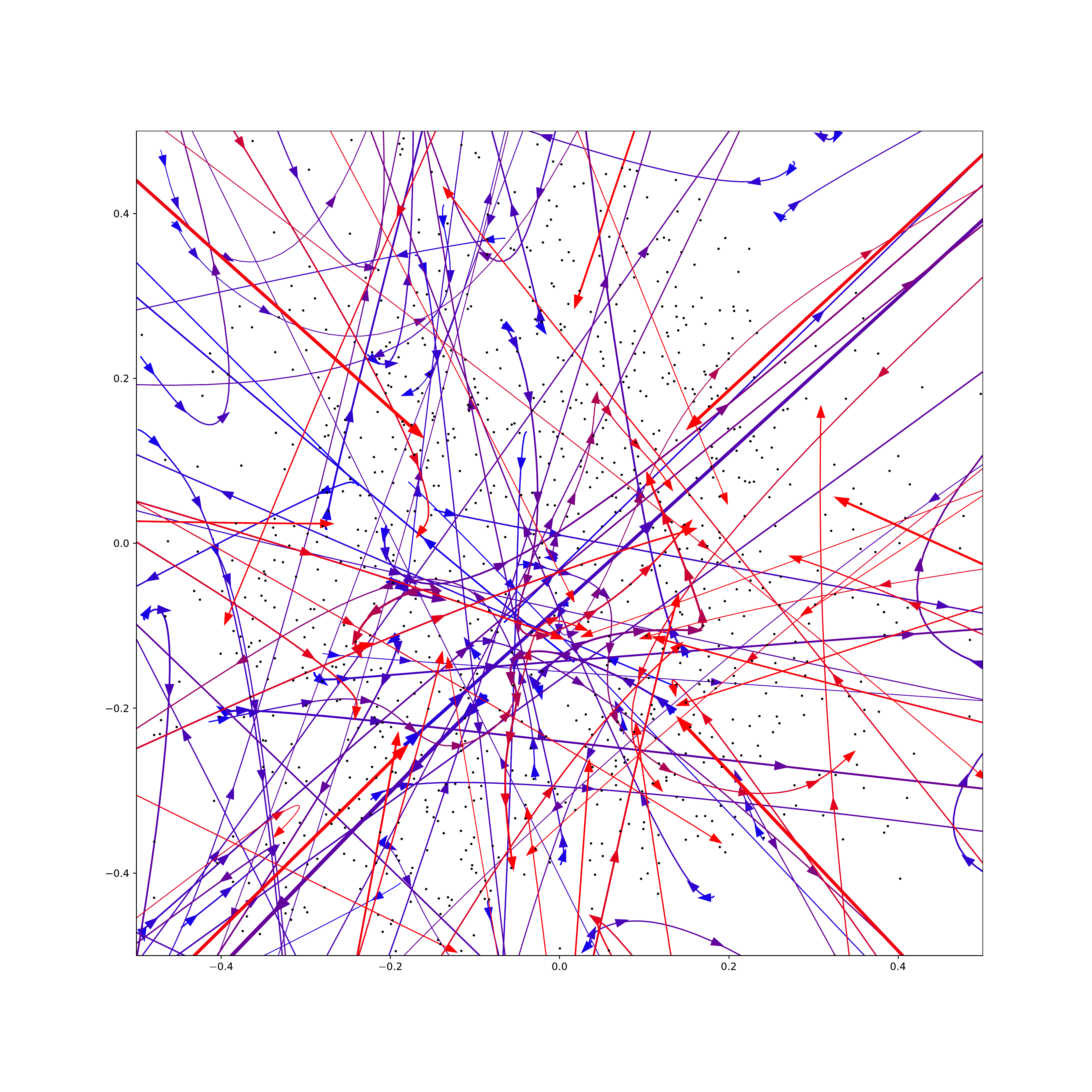}
    \caption{Clustering trajectories of editors. The trajectory size is proportional to the cluster population size. Largest trajectories contain approximately 3000 editors. The color progression from blue to red corresponds to the observed time. This figure presents both active editors and temporary editors. Active editors edit articles for a sustained period, highlighted by the color progression, which typically present a curved trajectory. They are characterized by a mid-size background and an important contribution to the articles. Alternatively temporary editors make fewer interactions, have a little background, and move faster with typically a straight line. }
    \label{fig:curved}
\end{figure}

\begin{figure}[t]
    \centering
    \includegraphics[width=\textwidth,height=\textheight,keepaspectratio]{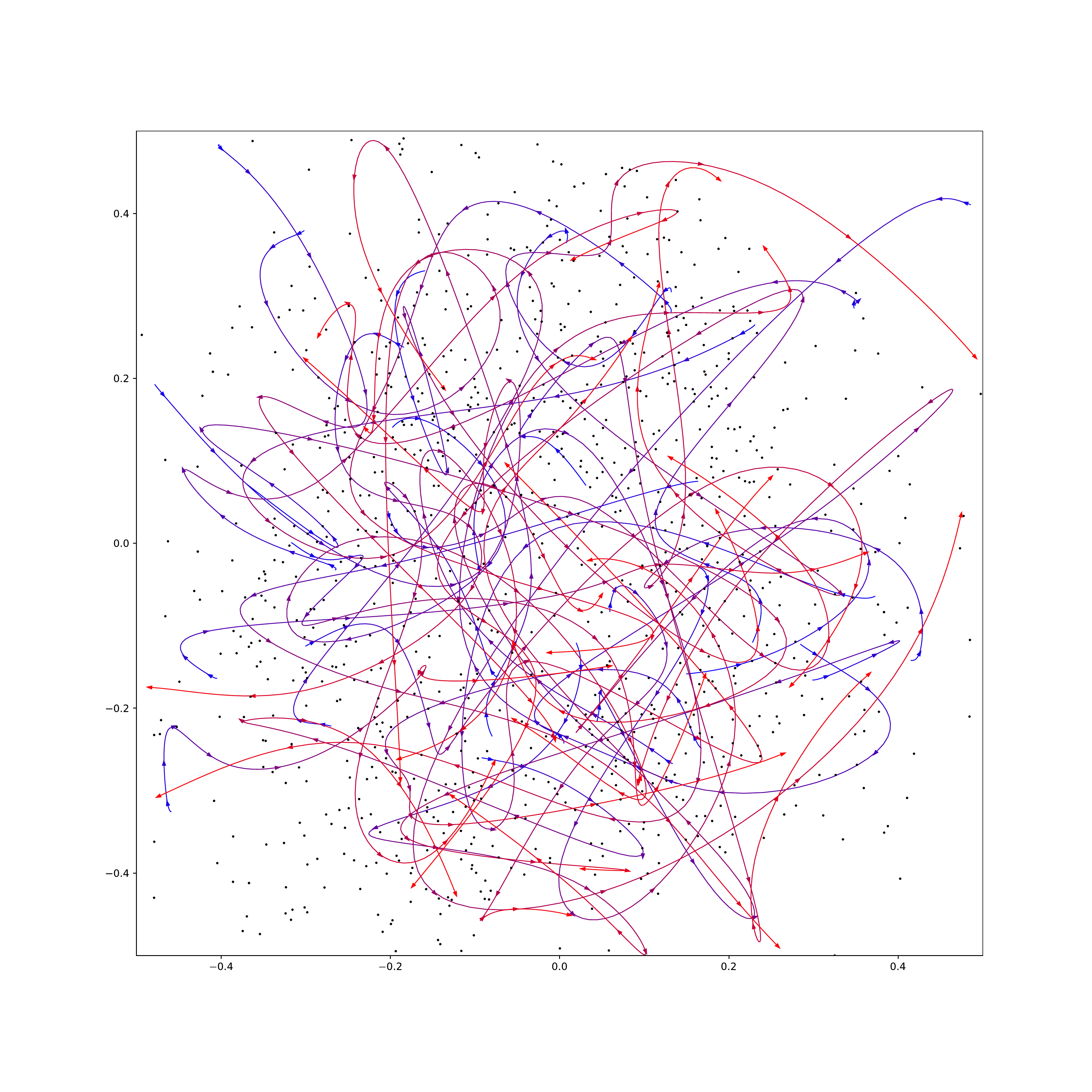}
    \caption{Independent trajectories: these editors do not have a cluster belonging. Their trajectory represents an independent behavior backed by strong expertise in their competence area. This figure presents only a subset of independent editors. We selected those with high centrality by filtering trajectories within [-0.5, 0.5]. The color progression from blue to red corresponds to the observed time.}
    \label{fig:independent}
\end{figure}

In the analysis three patterns can be identified among the editor trajectories: (1) independent editors with trajectories that explore a wide range of articles, see Figure~\ref{fig:independent}, (2)
 active editors who edited several articles with a possibly curved trajectory, see Figure~\ref{fig:curved}, and (3) temporary editors that entered, interacted, and exited using a straight trajectory.

\paragraph{Independent editors} These editors are likely highly experienced or specialized in various areas, as they edit a wide range of articles. They are difficult for others to replicate their latent patterns. These editors do not commonly belong to any cluster. They may be considered experts within their domain, and their contributions to Wikipedia may be highly valuable due to their depth of knowledge and expertise. They are shown mainly in Figure~\ref{fig:independent} although few examples are successfully clustered in Figure~\ref{fig:curved}.
\paragraph{Active editors} These editors enter or leave the cloud of articles with a possibly curved trajectory, see Figure~\ref{fig:curved}. They may be more casual or novice contributors who are focused on a specific set of articles. They may have a lower level of expertise than the editors with independent trajectories and may not engage with as many articles, but they can contribute valuable edits and improvements to the articles they interact with. 
\paragraph{Temporary editors} These editors enter, interact, and exit using straight trajectories. Some examples are shown in Figure~\ref{fig:curved} and more in detail in Figure~\ref{fig:straight}. These editors are considered snapshot editors. They make interactions only in a very short period of time. The arrows distance, which is approximately two years, highlights how fast these nodes move in the space. The proposed framework naturally model these trajectories as straight lines as no data supports a possible curvature outside the interaction interval.

Figure~\ref{fig:curved} and Figure~\ref{fig:independent} presents clusters and outliers respectively. These are obtained from the optimal radius selected by our method. The optimal radius hence captures a mixture of three behaviors. Alternatively the use of a sub-optimal radius can focuses on one single behavior. Figure~\ref{fig:straight} shows the clustering for a larger radius, capable of capturing the snapshot editors only. The usage of different radii, although sub-optimal from the model formulation perspective, can hence be informative. The fact that we need to use multiple radii to identify different behaviors in the trajectories may be due to the complexity and diversity of the data, as the editors' trajectories exhibit a wide range of behaviors and patterns.

\section{Conclusions}

The main contribution of this manuscript is the development of an efficient inference scheme for latent dynamic processes underlying a relational event process. The framework is general and can be extended to networks with weighted edges of any exponential family distribution, making it a useful tool for analyzing a wide range of data.
 
One key aspect of the model is the use of smooth spline functions to capture the latent trajectories of nodes in dynamic networks. This allows for a more accurate representation of the underlying dynamics than traditional static models. The model also employs a smoothness penalty for regulating the smoothness of the spline, and a clustering penalty for detecting shared trajectories among the nodes. This makes the model more interpretable and allows for the identification of patterns and behaviors within the network. The model can be run within the detected clusters and fit a nested latent space, which allows to reveal different levels of granularity of the relationships.

Another important aspect of the model is its scalability. The optimization is performed by the popular \textit{Adam} algorithm, which is not memory intensive and is very fast in computation. It can optimize nearly any function and learn the Hessian via the past gradients history. This allows the model to handle large networks with millions of nodes/parameters, which is going to be a common problem for future network analysis. Additionally, particular care has been given to handling sparse data and sparse parameter updates, which makes the model more robust.

The inference is conducted via Variational Bayes, finding an effective approximation of the posterior distribution for the complete set of parameters, including smoothness magnitude and clustering shrinkage.
Under the Variational formulation, the inference problems translates into a classic optimization problem, finding \textit{Adam} as a good ally.

This manuscript includes a simulation study that confirms the claims made in the manuscript, showing that the model behaves correctly in scenarios such as sparsity in the data, sparsity in the parameter update, clustering accuracy and consistency of the location estimator.

We applied the model to the Wikipedia complete edited page history. 
Differently to \cite{lerner2020reliability}, that analyzed this data with a 5 parameter model, our latent space model successfully fitted several millions of parameters. 
The application of the model to the Wikipedia data revealed various shared behaviors that are coherent with natural expectations. For example, some editors consistently modify Wikipedia pages over time, while others are more temporary editors. This differentiation between experts and non-experts shows that the model correctly identifies important behaviors in the Wikipedia editing patterns, which could help to understand the dynamics of the Wikipedia community and improve the quality of the articles.

Overall, the proposed model provides a powerful and interpretable tool for analyzing the dynamics of networks, and can help reveal the underlying patterns and behaviors of the nodes. The interpretability of the results makes it a valuable tool for understanding the underlying dynamics and making predictions about future behavior. This can be useful for a wide range of applications such as social network analysis, recommender systems and biological networks.  Given the popularity of a latent space representation in various emerging fields, possible extensions for our model include financial time series analysis, moving object detection, language generation and translation.

\backmatter

\section*{Declarations}
\bmhead{Funding}
The authors acknowledges funding from the Swiss National Science Foundation (SNSF 188534).

\bmhead{Competing interests}
The authors declare that they have no known competing financial interests or personal relationships that could have appeared to influence the work reported in this paper.

\bmhead{Availability of data and materials} The dataset used for the Wikipedia analysis is openly available in Wikipedia Edit Event Data 2018 (WikiEvent.2018) at \url{https://doi.org/10.5281/zenodo.1626323}.

\begin{appendices}

\section{a mini-batch cluster penalty for finite $\gamma_{\text{dist}}$}
\label{app:mbclust}
When we construct the mini-batch $B$, the chance of randomly sampling two close are almost zero for large networks. As a result, when $w_{ij}$ has relatively small radius only few elements are included in the kernel or, in case of a continuous kernel, have a sufficiently high weight. Therefore the vast majority of elements in the mini-batch are excluded. As a consequence, the level of sparsity for the gradient with respect to $c$ is even higher than for the splines. In order to make the gradient dense we propose to save the history of pairs that entered in the kernel at the previous iterations, then randomly sample $B^*$ in this set, where $ \mid B^* \mid =p$. The resulting mini-batch penalty is 
\begin{equation}
\label{eqn:conv_clustB_full}
P_{\text{clust}}^B = \gamma_{\text{aux}}\sum_{i=1}^p   \| \alpha_i - c_i \|^2 + \gamma_{\text{dist}}\sum_{i,j \in B^*} w_{ij} \| c_i - c_j \|^2.
\end{equation}
which has complexity linear in $p$. The mini-batch $B^*$ is sampled over the history of pairs for which $w_{ij}$ is positive. Notice that, similarly to the smoothness penalty, the full time sequence of the sampled nodes is included. 

\section{a discrete time model for sparse data}
\label{app:mixedsparsity}
We use this model formulation for the specific case when there exist dense connectivity within and sparse connectivity between communities. Hence one benefits from aggregating the data. We employ a non stratified case-control formulation of the Poisson likelihood \ref{eqn:lik-discN}. 
In order to obtain an unbiased estimate of the intercept, the likelihood term for the non-events need to be overweighted by $\frac{N_0}{n_0}$
\begin{equation}
\label{eqn:caseControlDiscrete}
 		\ell(\alpha) = \sum_{y_{t,ij}>0} \left[-\lambda_{i,j}(t) \Delta t + y_{i,j}(t) \log \lambda_{i,j}(t)\Delta t \right] + \frac{N_0}{n_0}\sum_{y_{t,ij}=0} -\lambda_{i,j}(t) \Delta t.
\end{equation}
Alternatively, the intercept absorbs the bias leading to the correct latent node positions.

\section{variational inference details}
\label{app:elbo}
In the Monte Carlo Variational approach some expectations can be solved analytically, leaving the Monte Carlo integration for those who are intractable.
Besides the non-tractable log-likelihood component, the remaining expectations can be solved as 
$$\mathbb{E}[P_{\text{smooth}}] = \frac{pd(k-1)}{2} \mathbb{E}[\log \gamma_{\text{smooth}}] - \mathbb{E}[\gamma_{\text{smooth}}] \sum_{i=1}^p\sum_{k=2}^m \mathbb{E}[ \left\|\alpha_{i,k} -\alpha_{i,k-1}\right\|^2] $$
$$\mathbb{E}[P_{\text{clust}}] = \frac{pdk}{2} \mathbb{E}[\log \gamma_{\text{clust}}] - \mathbb{E}[\gamma_{\text{clust}}] \sum_{i=1}^p \mathbb{E}[ \left\|\alpha_{i} - c_{i}\right\|^2] $$
where $(\log \gamma_{\text{smooth}}, \log \gamma_{\text{clust}})$ are Gaussian densities with  log-normal expectations $\mathbb{E}[\gamma_{\text{smooth}}] = e^{\mu_{\text{smooth}} + 0.5\sigma_{\text{smooth}}^2}$ and $\mathbb{E}[\gamma_{\text{clust}}] = e^{\mu_{\text{clust}} + 0.5\sigma_{\text{clust}}^2}$. The expectations of the normalizing constants are $\mathbb{E}[\log \gamma_{\text{smooth}}] = \mu_{\text{smooth}}$ and $\mathbb{E}[\log \gamma_{\text{clust}}]=\mu_{\text{clust}}$. The other expectations are simply functions of the first two moments $\mathbb{E}[\alpha_{i}] = \mu_i$ and $\mathbb{E}[\alpha_{i}^2] = \sigma^2_i + \mu_i^2$. 
The centroids $c_i$ can be safely held to be constant for various reasons. The first is that  $c_i$ is an averaging between several trajectories and hence its variance must be negligible compared to $\alpha_i$. The second is that $\mathbb{E}[ \left\|\alpha_{i} - c_{i}\right\|^2]$ is minimized by taking $c_i$ as degenerate, or nearly degenerate since that the prior would prevent the estimation of degenerate random variables.
The last remaining parameter $\gamma_{\text{radius}}$ cannot be updated by gradient, however is particularly easy to find a grid of candidate points from visual inspection of the latent space. We then select $\gamma_{\text{radius}}$ that maximize the lowerbound. 

The final component, $D_{\text{KL}}[q(\theta)  \| p(\theta)]$ can also find close form as done in the appendix of \cite{kingma2013auto}
$$ D_{\text{KL}}[q(\theta)  \| p(\theta)] = - \frac{1}{2} \sum_{i=1}^{p+2} 1 + \log\frac{\sigma_i^2}{\sigma_0^2} - \frac{\sigma_i^2}{\sigma_0^2} - \frac{(\mu_i - \mu_0)^2}{\sigma_0^2} $$
where $\mu_0, \sigma^2_0$ are respectively mean and variance of a Gaussian prior.

When running the Variational inference we might have the $\gamma_{\text{clust}}$ estimate being misleading for the case when both the number of cluster and the number of links are low. This because the penalty might become such big that is the major contributor to the lowerbound. The model hence prioritizes the minimization of  $\mathbb{E}[ \left\|\alpha_{i} - c_{i}\right\|^2]$ collapsing all trajectories and making $\gamma_{\text{clust}}$ unreasonably big. The lowerbound still make a correct clustering selection although some estimates of $\gamma_{\text{clust}}$ might not be coherent with the expectations, i.e. expecting $\gamma_{\text{clust}}$ low for low number of clusters. This behavior is paired by a substantial worsening in the likelihood, that reflects the introduction of the bias.

\begin{figure}
    \centering
    \includegraphics[width=\textwidth,height=\textheight,keepaspectratio]{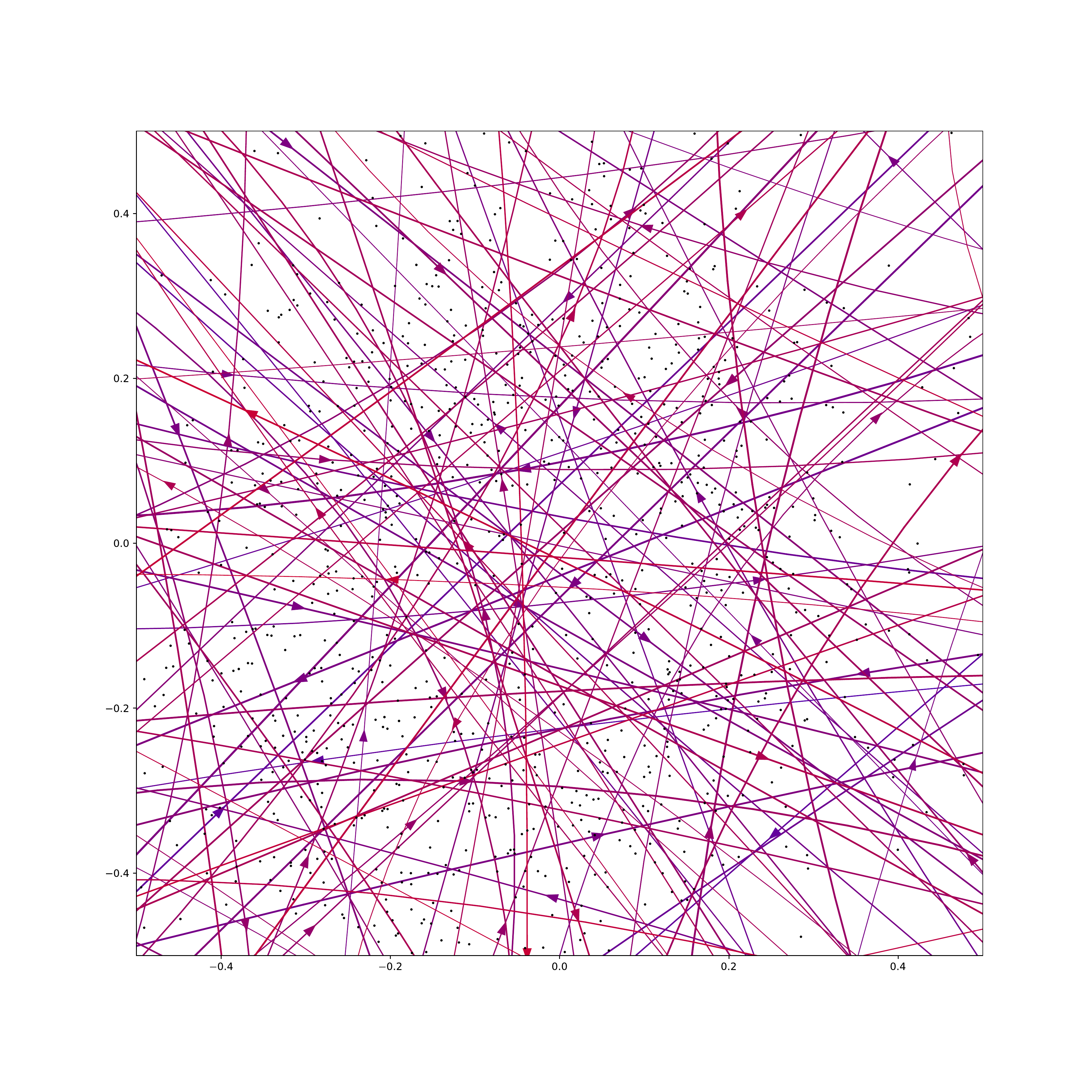}
    \caption{Snapshot editors. In-and-out editors active in a short period of time. Their competence area is little as they focus on few articles or topics. They typically have a straight trajectory. This figure presents the largest distance between the arrow heads, which is approximately two years.  This implies these editors are the fastest movers observed.  }
    \label{fig:straight}
\end{figure}

\end{appendices}

\bibliography{references}

\end{document}